\documentclass[]{spie} 

 \pdfoutput=1
\usepackage{amsmath,amsfonts,amssymb}
\usepackage{graphicx}
\usepackage[colorlinks=true, allcolors=blue]{hyperref}

\title{The adaptive optics lucky imager (AOLI): presentation, commissioning and AIV innovations}

\author[a,b]{Sergio Velasco}
\author[a,b]{Carlos Colodro-Conde}
\author[a,b]{Roberto L. L\'opez}
\author[a,b]{Alejandro Oscoz}
\author[c]{Lucas Labadie}
\author[a,b]{Yolanda Mart\'in-Hernando}
\author[d]{Antonio P\'erez Garrido}
\author[e]{Craig Mackay}
\author[a,b,f]{Rafael Rebolo}

\affil[a]{Instituto de Astrof\'isica de Canarias, c/ V\'ia L\'actea s/n, La Laguna, Tenerife E-38205, Spain.}
\affil[b]{Departamento de Astrof\'isica, Universidad de La Laguna, La Laguna, Spain.}
\affil[c]{I. Physikalsiches Institut, Universit\"at zu K\"oln, Z\"ulpicher Strasse 77, 50937 K\"oln, Germany}
\affil[d]{Universidad Polit\'ecnica de Cartagena, Campus Muralla del Mar, Cartagena, Murcia E-30202, Spain. }
\affil[e]{Institute of Astronomy, University of Cambridge, Madingley Road, Cambridge CB3 0HA, United Kingdom}
\affil[f]{Consejo Superior de Investigaciones Cient\'ificas, Madrid, Spain.}

\authorinfo{Further author information: (Send correspondence to S.V.)\\S.V.: E-mail: svelasco@iac.es}

\begin{document} 
\maketitle

\begin{abstract}
 Here we present the Adaptive Optics Lucky Imager (AOLI), a state-of-the-art instrument which makes use of two well proved techniques, Lucky Imaging (LI) and Adaptive Optics (AO), to deliver diffraction limited imaging at visible wavelengths, ~20 mas, from ground-based telescopes. Thanks to its revolutionary TP3-WFS, AOLI shall have the capability of using faint reference stars. In the extremely-big telescopes era, the combination of techniques and the development of new WFS systems seems the clue key for success. We give details of the integration and verification phases explaining the defiance that we have faced and the innovative and versatile solutions for each of its subsystems that we have developed, providing also very fresh results after its first fully-working observing run at the William Herschel Telescope (WHT). 
\end{abstract}


\section{INTRODUCTION}
\label{sec:intro} 

The Lucky Imaging (LI) technique, as suggested by \cite{1964JOSA} and named by \cite{1978JOSA}, was born as an alternative to AO to reach the diffraction limit in the optical bands. Images are taken at a very high frequency in order to select those intervals in which the atmosphere inside the collector tube through which the wavefront travels can be regarded as stable. If the best fraction of a bunch of images, those with smaller Strehl pattern, are stacked in a shift-and-add process, the equivalent to a high quality near-diffraction limit is obtained. The fraction of images that are selected for each target depends on the atmospheric conditions. The LI technique offers to ground-based telescopes an excellent and cheap method of reaching diffraction limited spatial resolution in the visible. 

One of the existing instruments to take advantage of the LI technique is FastCam, jointly developed by the Instituto de Astrof\'isica de Canarias (IAC) and the Universidad Polit\'ecnica de Cartagena (UPCT), described in \cite{2008SPIE}. FastCam, a common user instrument at the Carlos S\'anchez Telescope (CST, Teide Observatory, Canary Islands, Spain), routinely reaches the diffraction limit in the optical \textit{I} band both at the CST and the Nordic Optical Telescope (NOT, Roque de los Muchachos Observatory, Canary Islands, Spain), as on \cite{2011MNRAS}. In addition, FastCam has also obtained the image with the best resolution ever at the Canary Observatories (0.067” in \textit{I} band at the William Herschel Telescope, WHT, Roque de los Muchachos Observatory, Canary Islands, Spain), see \cite{2010SPIE}.

However, this technique suffers two important limitations: it can be applied only at telescopes with sizes below 2.5m, achieving a resolution similar to that of the HST, and most of the images are discarded, meaning that only relatively bright targets can be observed. 

AOLI is a state-of-the-art instrument conceived to beat these limitations by combining the two most successful techniques to obtain extremely high resolution, LI and Adaptive Optics (AO). This instrument is hence planned as a double system that includes an adaptive optics closed loop corrective system before the science part of the instrument, this last using LI. The addition of low order AO with a new Two Pupil Plane Positions wavefront sensor (TP3-WFS) \cite{Colodro} to the system before the LI camera enhances the reachable resolution as it removes the highest scale turbulence maximizing the LI process at larger telescopes. 

Aiming at this challenging goal we have built AOLI, see \cite{VelascoSEA} and \cite{CraigSPIE}, putting together the expertise of several institutions -IAC, IoA, UPCT, UC and ULL-, each group specialized in a different subject, corresponding to a part of the puzzle.

This instrument has seen several configurations in an important evolution since the very first design to the successful day of the commissioning, always trying to make it more compact and versatile. During the AIV process we found that a high accuracy in the optical alignment and the repeatability of the mechanical positioners were the clue to succeed. For this purpose, AOLI has been built in blocks over breadboards, which have been integrated and verified independently, each one completely removable and exchangeable. This “puzzle thinking” gives AOLI a high flexibility while keeping its strength allowing to improve and optimize each subsystem independently to mount the final puzzle.

The aim of AOLI is twofold: to provide extremely high spatial resolution and to offer a cheap and versatile instrument to be installed on different telescopes.

\begin{figure} [ht!]
\begin{center}
\includegraphics[height=5cm]{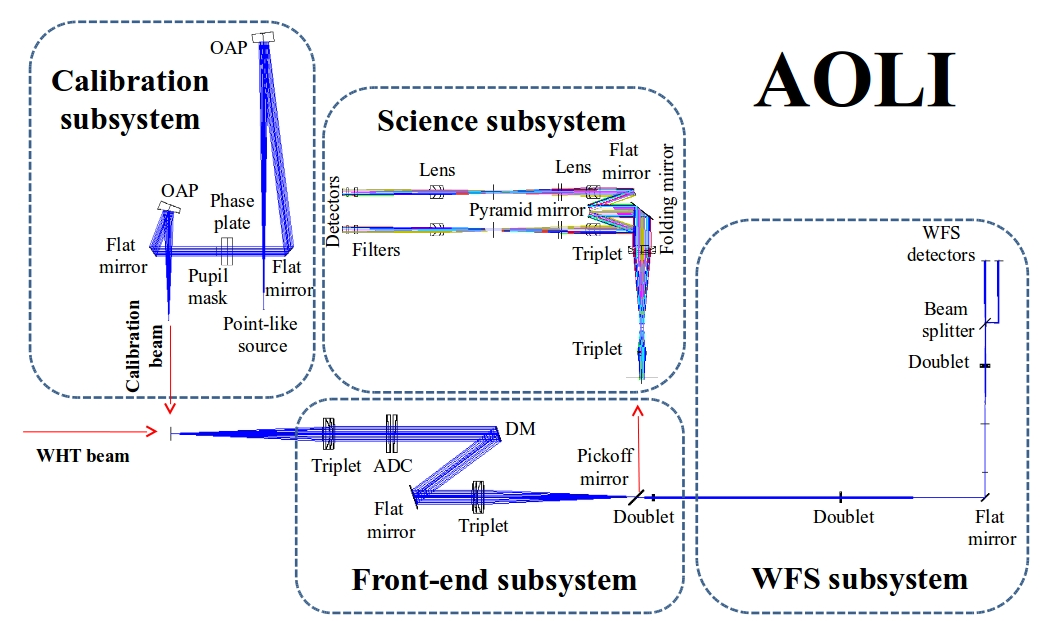}
\end{center}
\caption[layout] 
   { \label{fig:layout} 
AOLI's optical layout. The AO subsystem before the science camera rises the amount of images that can be used for lucky imaging, allowing the observation of fainter objects.}
\end{figure}

\begin{figure} [ht]
	\begin{center}
		\begin{tabular}{c} 
			\includegraphics[height=5cm]{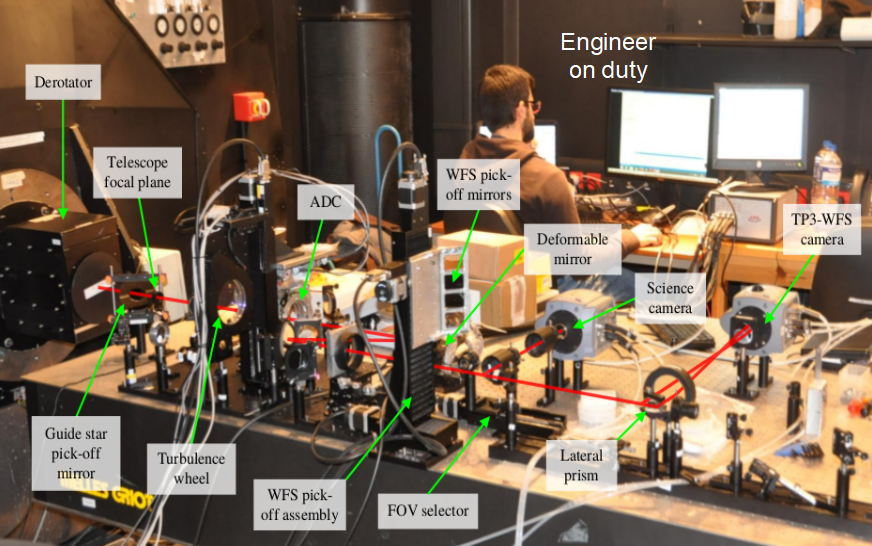}
			\end{tabular}
		\end{center}
		\caption[lay_etiq] 
		{ \label{fig:lay_etiq} 
		AOLI at the WHT with each o its subsystems mounted on a puzzle thinking way. The optical design has been optimized on the AIV phase, making AOLI a more versatile, lighter and simpler AO instrument.}
\end{figure} 

\section{AIV innovations}

To face the defiance that AOLI represents we have implemented a new philosophy of instrumental prototyping by modularizing all its components: simulator/calibrator, deformable mirror (DM), science and WFS modules. This modular concept \cite{RobertoSPIE} offers huge flexibility for changes, such as the addition of future developments and improvements or the hosting telescope. AOLI has now been restructured not only to make the AIV phase reliable but also to be able to integrate this system regarding different parameters (f-number, scale, WFS-type,) or to adapt it to different telescopes. AOLI, initially designed for the 4.2m William Herschel Telescope (WHT, Observatorio del Roque de los Muchachos, La Palma island, Spain) can be adapted to other telescopes,  including the 10.4m GTC (ORM, La Palma island).

The four  modules that compound the system are: 
\begin{itemize}
\item Telescope and turbulence simulator and calibrator \cite{MartaSPIE} (SimCal): it delivers a calibrated point-like source resembling the telescope f-ratio.
\item Science subsystem based on on EMCCDs with 3 different plate scales and FoV. 
\item WFS subsystem with a novel Two Pupil Plane Positions WFS (TP3-WFS), implemented for the first time on AOLI. It retrieves the wavefront by measuring the intensity of defocused pupil images taken at two planes.
\item Front-end + AO subsystem: 241 actuators ALPAO deformable mirror (DM) and conditioning optics.
\item Unique software allowing live-view processing (>2 million images per night)
\end{itemize}

\begin{figure} [ht!]
	\begin{center}
		\begin{tabular}{c} 
			\includegraphics[height=5cm]{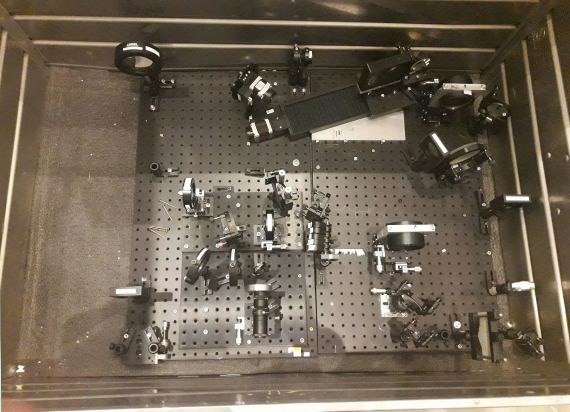}
			\end{tabular}
		\end{center}
		\caption[puzzle_box] 
		{ \label{fig:puzzle_box} 
			The SimCal puzzle piece aligned and ready to be transferred to the telescope from the AIV room.}
\end{figure} 

\begin{figure} [ht]
	\begin{center}
		\begin{tabular}{c} 
			\includegraphics[height=5cm]{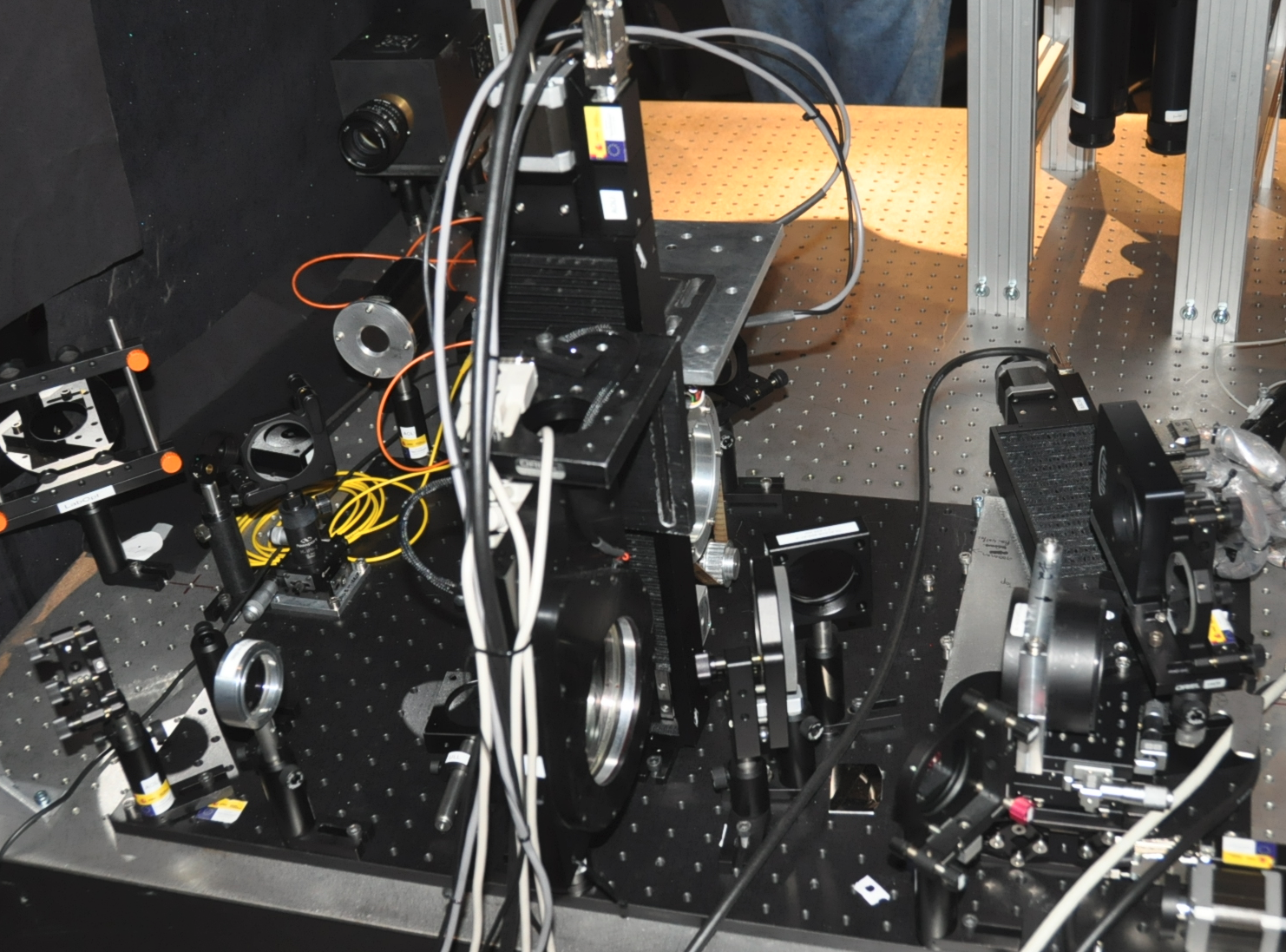}
			\end{tabular}
		\end{center}
		\caption[puzzle_wht] 
		{ \label{fig:puzzle_wht} 
			 The simulation and calibration system (SimCal) and the front-end subsystem merged in one puzzle piece.}
\end{figure}

\section{Commissioning}

On September 24th and 25th 2013, AOLI was installed at WHT's Nasmyth platform for a first commissioning to test its subsystems and overall efficiency without a fully developed AO subsystem. We were able to determine the plate scale (55.0$\pm$0.3 mas/pixel) and the PSF with a FWHM of 0.15 arcsec, probing them to be stable and to satisfy the specifications given. With just some seconds of on-sky integration we could probe the viability of the instrument offering some scientific results, see \cite{Velasco2016} and \cite{VelascoSEA16}.

\begin{figure} [ht!]
	\begin{center}
		\begin{tabular}{c} 
			\includegraphics[height=3cm]{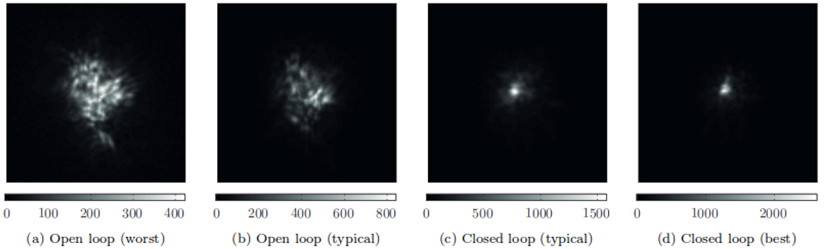}
			\end{tabular}
		\end{center}
		\caption[hip1006] 
		{ \label{fig:hip1006} 
			Images of HIP10644 with and without AO corrections, showing worst, typical and best images according to the maximum pixel value. It seems clear that the closure of the control loop has a positive impact on the quality of the acquired images.}
\end{figure}

The first commissioning of the full instrument, including a fully working AO, took place at WHT on May 21st, 2016. The weather conditions that night were not optimal for high spatial resolution techniques due to a seeing around 2.2 arcsec, well above what is generally taken as a limit for AO systems, and full Moon. Despite of it, we closed the loop (see fig. \ref{fig:hip1006})  with different magnitudes stars making use of the TP3-WFS and an ALPAO deformable mirror with 244 actuators. In addition, the system was fed with a calibration made in real time from the image of the pupil from the reference star itself.

\section{Closing the loop on extended sources}

\begin{figure} [ht!]
	\begin{center}
		\begin{tabular}{c} 
			\includegraphics[height=5cm]{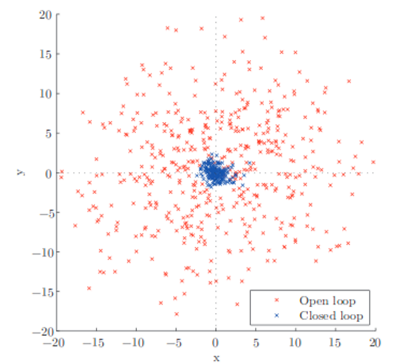}
			\end{tabular}
		\end{center}
		\caption[neptune] 
		{ \label{fig:neptune} 
			Coordinates of Neptune over the science camera. The dispersion of such coordinates is clearly reduced when applying AO with the tip-tilt control activated.}
\end{figure}

With the TP3-WFS we have achieved to close the loop on extended sources, such as Neptune, (see fig. \ref{fig:neptune})

\section{CONCLUSIONS AND FURTHER WORK}

As a consequence of its modularity, AOLI@WHT is much smaller and more efficient than its first design. This success has lead us to condense it to get a portable, versatile and easy to integrate system, ALIOLI (Adaptive and Lucky Imaging Optics Lightweight Instrument) \cite{Velasco2018b}.

\section{MULTIMEDIA FIGURES}

You can see some images, animations and new data over this instrument in \href{http://www.iac.es/proyecto/AOLI/}{IAC AOLI project homepage} [\url{http://www.iac.es/proyecto/AOLI/}]

\acknowledgments 

This paper is based on observations made with the William Herschel Telescope operated on the island of La Palma by the Isaac Newton Group in the Spanish Observatorio del Roque de los Muchachos of the Instituto de Astrof\'isica de Canarias

\bibliography{main} 
\bibliographystyle{spiebib} 

\end{document}